\documentclass[prl,aps,twocolumn,preprintnumbers,superscriptaddress]{revtex4}
\usepackage{latexsym,epsfig,amssymb,amsfonts,amsmath,graphicx,color,bm,times,amstext,epstopdf,braket,float,ulem}
\usepackage{bbm}

\begin{document}

\title{Correlation-driven Lifshitz transition at the emergence of the pseudogap phase in the two-dimensional Hubbard model}
%for cuprates}

\author{Helena Bragan\c{c}a}
\affiliation{Departamento de F\'isica, Universidade Federal de Minas Gerais,
C. P. 702, 30123-970, Belo Horizonte, MG, Brazil} 
\affiliation{Laboratoire  de  Physique  des  Solides,  Univ.   Paris-Sud,
Universit\'e  Paris-Saclay,  CNRS  UMR  8502,  F-91405  Orsay  Cedex,  France}
\author{Shiro Sakai}
\affiliation{Center for Emergent Matter Science, RIKEN, Wako, Saitama 351-0198, Japan}
\author{M. C. O. Aguiar}
\affiliation{Departamento de F\'isica, Universidade Federal de Minas Gerais,
C. P. 702, 30123-970, Belo Horizonte, MG, Brazil}
\author{Marcello Civelli}
\affiliation{Laboratoire  de  Physique  des  Solides,  Univ.   Paris-Sud,
Universit\'e  Paris-Saclay,  CNRS  UMR  8502,  F-91405  Orsay  Cedex,  France}

\date{\today}

\begin{abstract}
We study the relationship between the pseudogap and Fermi-surface topology in the two-dimensional Hubbard model 
by means of the cellular dynamical mean-field theory.
%We perform a cellular dynamical mean field study of the paramagnetic state of the two-dimensional Hubbard model.
We find two possible mean-field metallic solutions on a broad range of interaction, doping and frustration: a 
conventional renormalized metal and an unconventional pseudogap metal. At half-filling, 
the conventional metal is more stable and displays an interaction-driven Mott metal-insulator transition.
However, for large interaction and small doping, region that is relevant for cuprates, 
the pseudogap phase becomes the ground state. 
By increasing doping, we show that a first-order transition from the pseudogap to the conventional metal is tight to
a change of the Fermi surface from hole to electron like, unveiling a correlation-driven mechanism for a Lifshitz transition. 
This explains the puzzling link between pseudogap phase and Fermi surface topology which has been pointed out in recent 
experiments.      
\end{abstract}

\maketitle

\begin{figure*}[t]
  \begin{center}
     \includegraphics[width=0.8\linewidth]{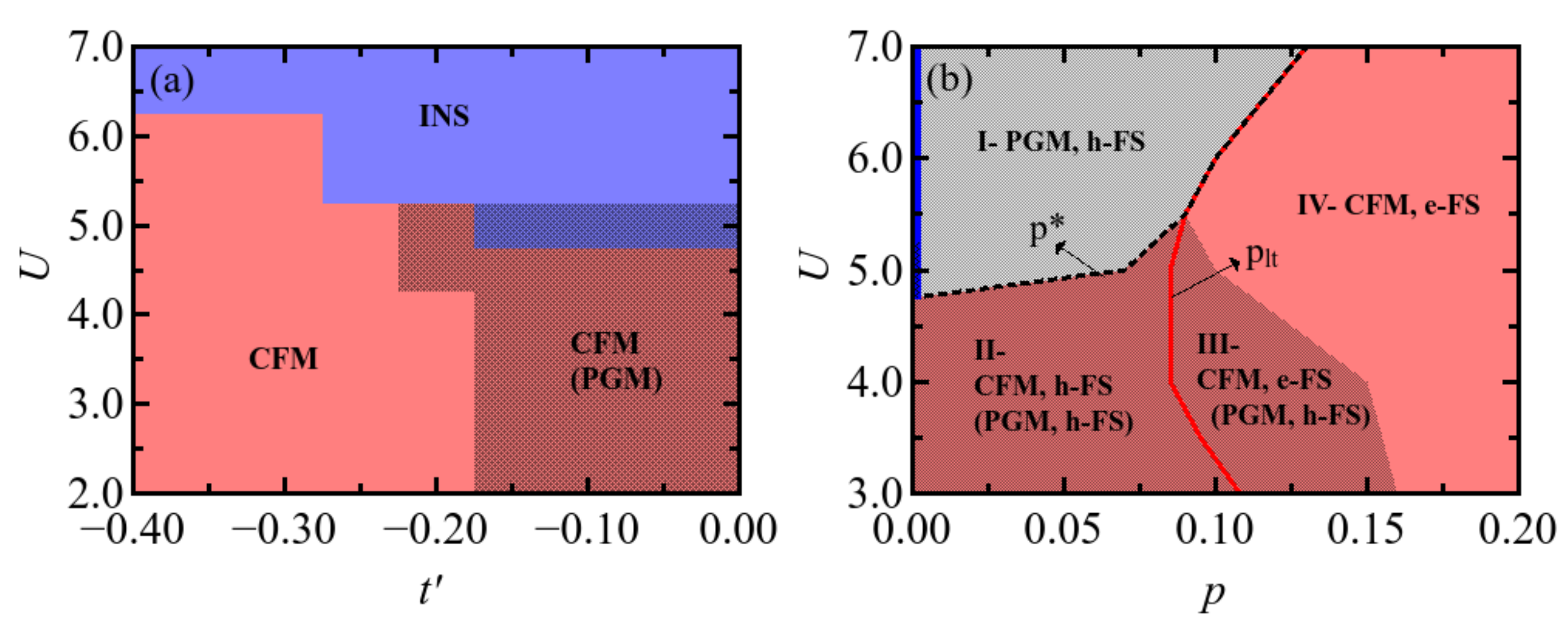}
\caption{ (a) $U- t^{\prime}$  phase diagram for the normal state of the two-dimensional Hubbard model at half-filling ($p=0$). 
  The blue, pink and dark regions bound respectively a Mott insulator (INS), a correlated Fermi-liquid metal (CFM) and
  a coexisting pseudogap metal (PGM). The CFM has always lower energy than the PGM (in parenthesis). 
  (b) $U-p$ phase diagram with $t^{\prime}=-0.1$. The blue vertical line at $p=0$ is the INS.
  The continuous red line $p_{lt}$ marks a Lifshitz transition, where the FS topology changes from electron-like (e-FS) to
  hole-like (h-FS). The dashed black line $p^{*}$ marks the boundary where the PGM energy becomes higher than the CFM one. 
%\blue{The blue horizontal line on the left indicates the Mott insulator phase, found at zero doping.} 
  %which can be identified as the PG endpoint $p^{*}$ of the cuprates. As in (a), in the black hatched region
  %the CFM and the (unstable) PGM co-exist.} 
    }\vspace{-0.5cm} 
    \label{HF1}
  \end{center}
\end{figure*}

In order to understand superconductivity~\cite{bcs,migdal58,eli60} one must first understand 
the normal metallic state, appearing above a critical temperature ($T_c$), 
from which it takes its roots. The high-$T_c$ superconductivity in cuprates remains unsolved, 
mainly because its normal metallic state, the pseudogap (PG) phase, 
has not been well understood.
It has been therefore a central issue to establish the origin of the pseudogap and its
relation with the high-$T_c$ superconducting mechanism\cite{Norman05}. 
%reveals as much puzzling 
%as the high-$T_c$ mechanism itself~\cite{Norman05}. 
The PG has been revealed~\cite{alloul89,warren89} in spectroscopic
responses~\cite{damascelli03} and thermodynamic and transport properties~\cite{timusk99}
by a loss of spectral weight, which departs from 
the conventional Fermi liquid (FL) theory of metals~\cite{abrikosov63,nozieres64}. 
Recent experiments have pointed out that when a Lifshitz transition  
[i.e. a change of Fermi Surface (FS) topology from electron-like (e) 
to hole-like (h)] is tuned in the PG phase of a cuprate material, 
the PG ends  abruptly. This takes place for instance at a doping 
$p^{*}$ on the overdoped region of Bi$_2$Sr$_2$CuO$_{6+\delta}$~\cite{piriou11},
Bi$_2$Sr$_2$CaCu$_2$O$_{8+\delta}$~\cite{benhabib15,loret17} and in 
La$_{1.6-x}$Nd$_{0.4}$Sr$_x$CuO$_4$\cite{doiron17}. 
%upon application of hydrostatic 
%pressure. 
This finding has been strongly debated, as it 
challenges our current understanding of the PG phase\cite{Norman05} 
and its relation with superconductivity

%why and 
%on which region of the phase diagram the PG emerges.
%Spectroscopic experiments on monolayer Bi$_2$Sr$_2$CuO$_{6+\delta}$~\cite{piriou11} and on 
%bilayer Bi$_2$Sr$_2$CaCu$_2$O$_{8+\delta}$~\cite{benhabib15,loret17} have revealed that the PG 
%ends abruptly at a doping $p^{*}$ on the overdoped region, which 
%appears close to a Lifshitz-transition, 
%where the Fermi surface (FS) changes its character from electron-like (e) to hole-like (h), 
%suggesting a tight relation between the PG and the FS topology. This
%link has been recently confirmed by hydrostatic pressure experiments 
%on La$_{1.6-x}$Nd$_{0.4}$Sr$_x$CuO$_4$\cite{doiron17}.
%on \gr{La$_{2-x}$Sr$_x$CuO$_4$ [SS:Please verify]}~\cite{doiron17}, 
%which have shown that $p^*$ and $p_{lt}$ tightly move together upon the application of 
%hydrostatic pressure.

Here we give a rational explanation to all these observations within the framework of the two-dimensional
Hubbard model solved with the cellular dynamical mean field theory (CDMFT)~\cite{Maier05,CDMFT,AMT06}. 
We first show that two metallic solutions exist:
a rather regular correlated Fermi-liquid metal (CFM), and a PG metal (PGM), which violates Fermi 
liquid theory, by developing a pole-divergence in the self-energy. 
%This gives origin to the PG in the spectral density. 
This result could account for contradicting reports about the existence of the 
%correlation-driven 
Mott metal-insulator transition (MIT) at half-filling (zero doping) in two dimensions. 
The PGM is metastable at weak interactions, having higher energy than the CFM. However, by increasing interaction 
at low doping (region relevant for underdoped cuprates) the PGM emerges as the stable phase, 
up to the doping value $p^{*}$. 
This is consistent with the CDMFT results of Sordi et al.\cite{AMT10}.
%, which is the model equivalent of the PG ending point in the cuprates.
Most importantly, we show that the PGM is bound to have always a h-FS. 
The CFM instead can undergo a Lifshitz transition 
%from a h-FS to an e-FS 
at a doping $ p_{lt}$. 
However, for strong interaction the CFM is stable only for doping $p> p_{lt}$, i.e. it has always an e-FS.
%and in the doping range where the CFM is the stable solution, the FS is always electron-like. 
Hence the transition from the PGM to the CFM at $p^{*}$ is accompanied by a corresponding change from a h-FS to an e-FS,
unveiling a novel correlation-driven mechanism of the Lifshitz transition.
These results explain %the experimental puzzle described above, namely
why the PG must sharply end when a Lifshitz transition occurs\cite{piriou11}, 
or is tuned by pressure\cite{doiron17}, in cuprates.
  
%why the PG ending point in cuprates is tightly bound to a change of FS topology and  
%why this appears to be a first-order one.
%%%%%%%%%%%%%%%%%%%%%%%%%%%%%%%%%%%%

We consider the two-dimensional one-band Hubbard model: % given by the Hamiltonian
\begin{eqnarray} 
 \mathcal{H}=\, - \sum_{ \mathbf{k}  \sigma} \, \xi_{\mathbf{k}} \, c^\dagger_{\mathbf{k}\sigma}c_{\mathbf{k} \sigma}  
%-t' \sum_{\langle \langle ij \rangle \rangle \sigma} \left( c^\dagger_{i\sigma}c_{j\sigma} + h.c.\right) \nonumber \\
% &+& 
+ U \sum_{i } n_{i \uparrow}n_{i \downarrow},
%- \mu \sum_n n_i. 
\label{HubbardModel}
\end{eqnarray}
where 
$c_{\mathbf{k} \sigma}= (1/\sqrt{L}) \sum_{i} \exp(-\textit{i} \mathbf{k} \cdot \mathbf{r}_i) \, c_{i,\sigma} $ 
destroys an electron with spin $\sigma$ and momentum $\mathbf{k}$,  
$n_{i \sigma}=c^\dagger_{i \sigma}c_{i \sigma}$ is the density operator on site $i$ of 
a $L$-site square lattice. 
%the electronic dispersion is given by
$\xi_{\mathbf{k}}= -2t (\cos k_x + \cos k_y)- 4t^{\prime} \cos k_x \cos k_y- \mu$,
where $t$ ($t^{\prime}$) is (next) nearest-neighbor-site hopping integral, $\mu$ the chemical
potential controlling the doping level $p= 1- (1/L) \sum_{i,\sigma} \langle n_{i\sigma} \rangle$. 
We implement the CDMFT at zero temperature ($T=0$) using Lanczos. This maps $\mathcal{H}$ onto a 2$\times$2 cluster coupled to an 
8-site bath~\cite{Marcello2009,Capone2004,TeseMarcello} [see section I of the Supplemental Material (SM)]. %for a detailed description of the impurity solver]. 
The numerical calculation provides the frequency dependent Green's function
in the corner points
%symmetric points $\mathbf{k}=(0, \pi), \ (\pi, 0), \ (\pi,\pi)$ and $(0,0)$ 
of the first quadrant of the Brillouin zone (BZ). 
To obtain the lattice quantities in momentum space we perform a periodization based on the cumulant~\cite{PM,PM2,SIZE}. 
We calculate the total energy as described in 
Ref.~\cite{RefEnerg} and in section II of the SM, which includes
Ref.~\cite{Mahan}.  
We set $t=1$ %as the unit of energy 
and explore the paramagnetic phase diagram in the $U-t^{\prime}$ space at half-filling %(undoped system $p=0$) 
(Fig. \ref{HF1}a), and the $U-p$ space at fixed $t^{\prime}=-0.1$ (Fig. \ref{HF1}b).   
At $T=0$, the ground state is broken symmetry phase: antiferromagnetism
at half-filling and small doping, and superconductivity upon doping. These phases 
have been widely studied within CDMFT ~\cite{Maier05,BGK06,AMT06,SSK08,ferrero09,AMT10, Gull10,Gull13,Gull15}.
Here we focus on the paramagnetic solution which, albeit being the normal-state ground-state 
only at $T> T_c$, is the mean-field phase from which broken orders take roots. 
This allows us to study the FS topology and its relation with the pseudogap.

%CDMFT well captures the phases appearing on the cuprate phase diagram,
%including the Mott insulator, the antiferromagnetism, the PG state and the 
%superconductivity
%including normal and superconducting states~\cite{Maier05,BGK06,AMT06,SSK08,ferrero09,AMT10, Gull10,Gull13,Gull15}. 
%\gout{The theoretical $p$ values are however smaller than the typical experimental dopings,
%hence we limit ourself to qualitative comparisons, focusing on
%the paramagnetic solution at zero temperature.}} 

%%%%%%%%%%%%%%%%%%%%%%%%%%%%%%%%%%%%%%%%%%%%%%%%%%%%%%%%%%

%
%In this case the ground state is an antiferromagnetic insulator~\cite{Hirsch85,White89}, 
%with the N\'eel temperature $T=0$ in two 
%dimensions %, in accordance with the Mermin-Wagner theorem~
%\cite{MerminWagner}. 
%One can however avoid the antiferromagnetism by 
%imposing a paramagnetic mean-field solution.  

We start with the half-filled system [Fig.~\ref{HF1}(a)].
A relevant question is whether in two dimensions a gap 
is present in the paramagnetic solution at any small $U$, 
like in one dimension~\cite{Lieb68} and as it was proposed 
by P.W. Anderson~\cite{Anderson97}, or whether the system
becomes a regular metal under a critical $U_c$, i.e. there is a Mott MIT, 
like in infinite dimensions\cite{DMFT}.
%%%%%%%%%%%%%
This issue has been considered by various groups using quantum cluster 
methods~\cite{Jarrell,Imada,PHK,Tremblay,Toschi,ATM17}, but it has not been completely clarified. 
In these studies it was considered the particle-hole symmetric $t^{\prime}=0$ case, 
which is especially singular because a $\mathbf{k}= (\pi,\pi)$ nesting vector acts 
on the whole FS producing divergent susceptibilities. It is very likely then that 
at $T=0$ a gap always opens in the system. To verify 
Anderson's conjecture we have considered $t^{\prime} \neq 0$.
%%%%%%%%%%%%%%%%
Our main result is that, for $U<U_c$, %less than the critical value, 
we find two different metallic solutions, the CFM and the PGM,
%one is a Fermi liquid (CFM), the other a non-Fermi liquid (PGM), 
as it is shown in Fig.~\ref{HF1}(a).
%on the $U$-$t^{\prime}$ phase diagram of Fig.~\ref{HF1}(a). 
The PGM coexists with the CFM for a broad range of $U$ and 
$t'$ values, disappearing only for large %frustrating hopping parameter 
$|t^{\prime}|$. 
For interaction greater than $U \simeq 5t$ we recover 
the well known Mott insulating phase. 
We shall now show that the CFM is the FL solution displaying the Mott transition, in agreement with 
the statements of publications~\cite{Imada,PHK,Tremblay,ATM17}, while the PGM solution presents
always a gap in the spectra %and appears as the continuation at weak interactions $U$ of the Mott insulating state, 
reminiscent of the solution found in the works of Ref.~\cite{Jarrell,Toschi}. 
%%%%%%%%%%%%%%%%%%
\begin{figure}[h]
 \begin{center}
  \includegraphics[width=\linewidth ]{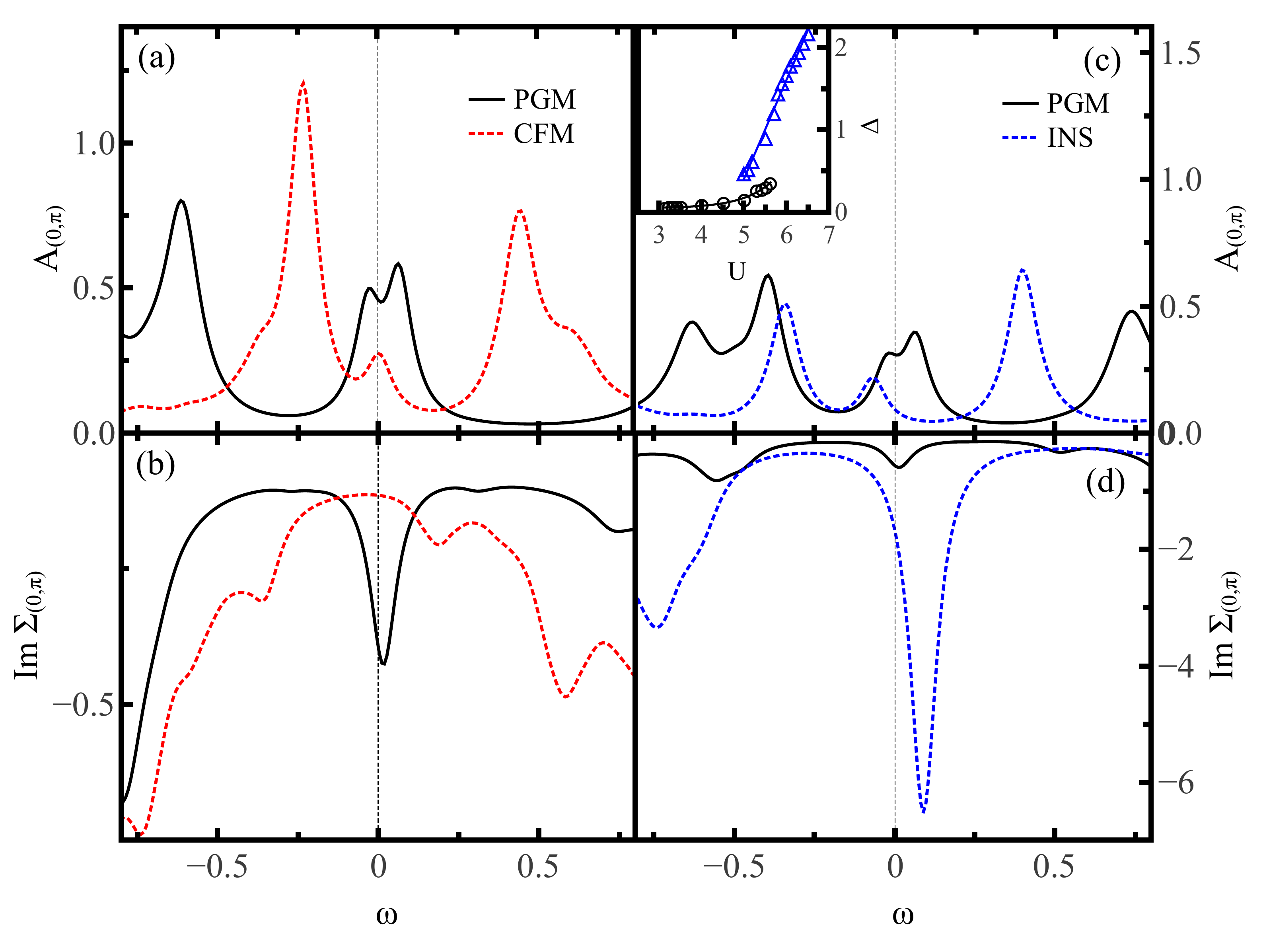}
 \caption{ Low-energy spectral function $A(\mathbf{k},\omega)$ (a) at $\mathbf{k}=(0,\pi)$
           and corresponding imaginary part of the self-energy Im$\Sigma(\mathbf{k},\omega)$ (b)
           of the two coexisting CFM and PGM at half-filling, $U=4.0$, $t^{\prime}=-0.1$. 
           Comparison of $A(\mathbf{k},\omega)$ (c) and Im$\Sigma(\mathbf{k},\omega)$ (d) between 
           the PGM and the Mott insulator for $U=5.0$, $t^{\prime}=-0.1$. 
           Inset: spectral gap of $A(\mathbf{k},\omega)$ as 
           a function of $U$ for the PGM (circles) and the Mott insulator (triangles).
%\gr{[SS:Please correct y labels of the figures 2 and 4, too.]}
}
 \label{HF2}
 \end{center}
 \end{figure} 

To this purpose, we set $t^{\prime}= - 0.1$ and 
display in Fig. \ref{HF2} the spectral function $A(\mathbf{k},\omega)=$ - $\frac{1}{\pi}$Im$G(\mathbf{k},\omega)$
%\gr{[SS:Please verify]}} 
and the imaginary part of the self-energy Im$\Sigma(\mathbf{k},\omega)$
at $\mathbf{k}=(0,\pi)$, close to the Fermi level ($|\omega|< 0.8 t$)
(see SM section III for a broader $\omega$-range %the plots of $A(\mathbf{k},\omega)$ in a broader range of $\omega$,
including the Hubbard bands). 
The CFM (red-dotted curve) displays typical features of a 
FL: finite spectral peak at the Fermi level $\omega=0$ (Fig. \ref{HF2}.a) and $\sim \omega^2$ behavior in
Im$\Sigma$ (Fig. \ref{HF2}.b).
%the imaginary part of the self-energy  
%On the other hand, 
The PGM (black-solid curve) displays sharply 
distinct features. $A(\mathbf{k},\omega)$ has a minimum at $\omega=0$ (Fig. \ref{HF2}.a), showing a small gap $\Delta$, 
which we plot in the inset of Fig. \ref{HF2}(c) (circles) as a function of $U$ together with the insulator gap (triangles).
Im$\Sigma$ displays a pole-like divergence (Fig. \ref{HF2}.b), which breaks the FL. 
This behavior of the self-energy is similar to what is expected in a Mott insulator (blue-dotted curve in Fig. \ref{HF2}.d), 
whose gap is always characterized by a pole in the self-energy, though in the PGM the intensity of the divergence is 
reduced and the two solutions are not smoothly connected, showing a coexistence region (inset of Fig. \ref{HF2}.c).
%%%%%%%%%%%%%%%%%%%%%%%%%%%%%%%%%

\begin{figure}[h]
  \begin{center}
    \includegraphics[width=\linewidth]{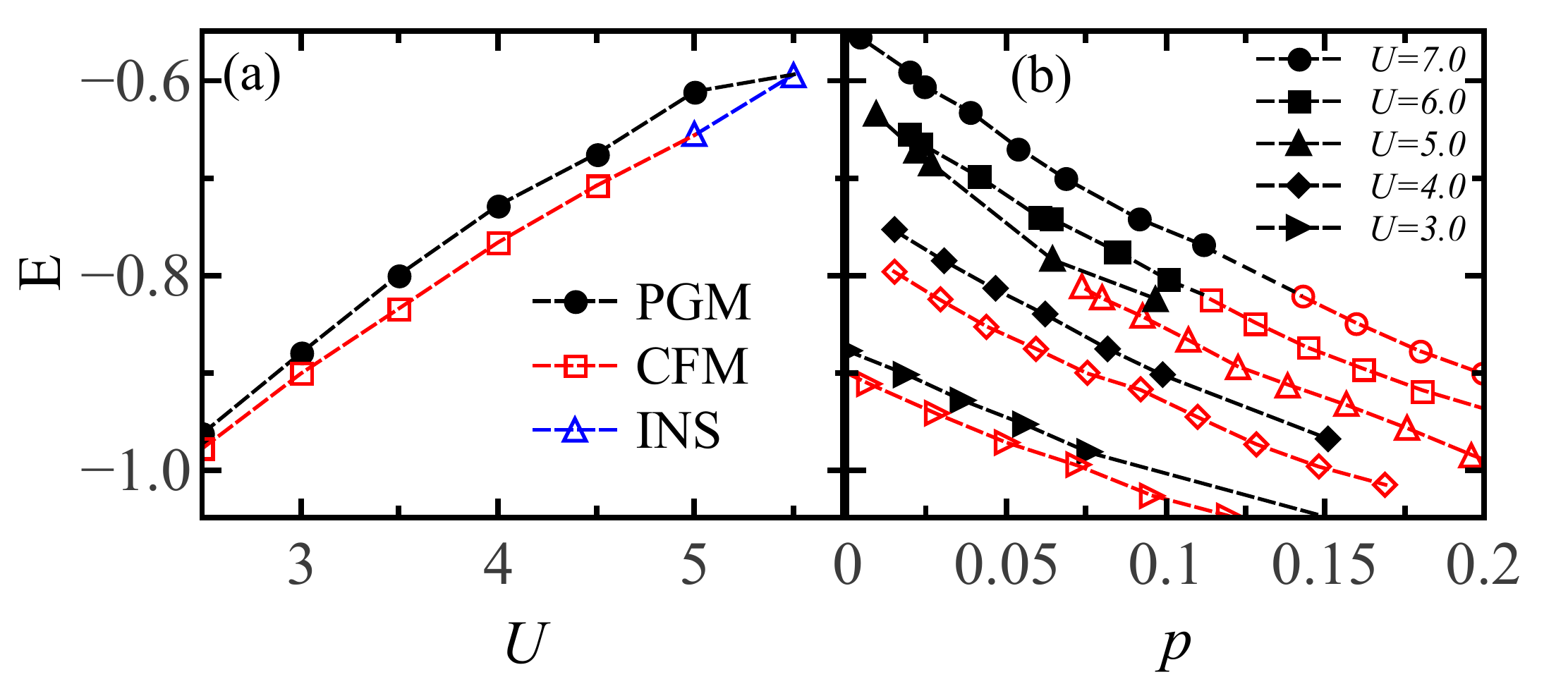}
    \caption{Total energy E (a) as a function of interaction at half-filling and (b) as a function of doping, both for $t^{\prime} = -0.1$.
Red-open symbols correspond to the CFM solution, while black-filled ones represent the PGM. } 
    \label{HF3}
  \end{center}
\end{figure}

We find that at half-filling and different $t^{\prime}$ %, within our CDMFT \gr{solutions}, 
the CFM always shows a lower energy than the PGM, as displayed in Fig. \ref{HF3}(a). This validates the concept of
Mott MIT also in the two-dimensional Hubbard model (see SM for the kinetic and potential energies).
%It is useful to separate the potential and kinetic energy contributions to the total energy (Fig. \ref{HF3}.b,c): 
%we observe that the PGM has a higher double occupation of sites $D$, i.e. a higher potential energy $U D$, 
%and a lower (negative) kinetic energy $K$ than the 
%CFM. This is reminiscent of the Anderson's Resonating Valence Bond (RVB) solution to the Heisenberg model~\cite{Anderson97}, 
%which favours a gain in kinetic energy with respect to the paramagnetic solution~\cite{Tocchio}. 
The unstable PGM remains however reminiscent of Anderson's RVB theory~\cite{Anderson97,Tocchio}, though relevant differences
have been already pointed out~\cite{Shiro}. 
%%%%%%%%%%%%%%%%%%%%%%%%%%%%%%%

%Even though the PGM is higher in energy than the CFM, 
The PGM can become however the relevant lowest-energy phase 
upon doping. 
%and by increasing correlation ($U \ge 4.5t$), as we show in Fig.~\ref{HF1}(b), where we keep 
%the particle-hole-breaking parameter 
%$t^{\prime}=-0.1$ fixed. 
%This is indeed the region corresponding to the underdoped cuprates. 
In the $U$-$p$ phase diagram of Fig.~\ref{HF1}(b), where $t^{\prime}=-0.1$, 
we can clearly identify three  regions: The PGM (I), the CFM (IV) and %, at weak correlation, 
a coexisting CFM-PGM region (II and III). %, like for half-filling.
In this latter case (see Fig.~\ref{HF3}.b for $U=3$, $4$), %\gout{where we display the total energies of the system,} 
%when the two solutions co-exist, 
the CFM has always the lowest total energy. 
For $U > U_{c} \simeq 5 t$, however, at small doping (region relevant for underdoped cuprates)
the PGM emerges as the stable solution, while the CFM persists at high dopings.  
These results are consistent with those of ref.~\cite{AMT10,Faye17}  
which shows a small first-order coexistence region
between regions I and IV, %with a continuous-time quantum Monte Carlo \gout{(CT-QMC)} 
%implementation of the CDMFT. 
which closes at a tricritical point and become continuous for $U \ge 7t$. 
We think that our Lanczos-implementation has difficulties to enter into this small coexistence region,
and we cannot say if the transition becomes second order for $U \ge 7t$.  %than the one considered. 
However the first-order character of the transition in the region that we considered is shown 
by the fact that the PGM and the CFM are not smoothly 
connected. %, as shown by  continuing them to weak interactions $U$. 

%%%%%%%%%%%%%%%%%%%%%%%%%%%%

\begin{figure}[h]
  \begin{center}
    \includegraphics[width=\linewidth]{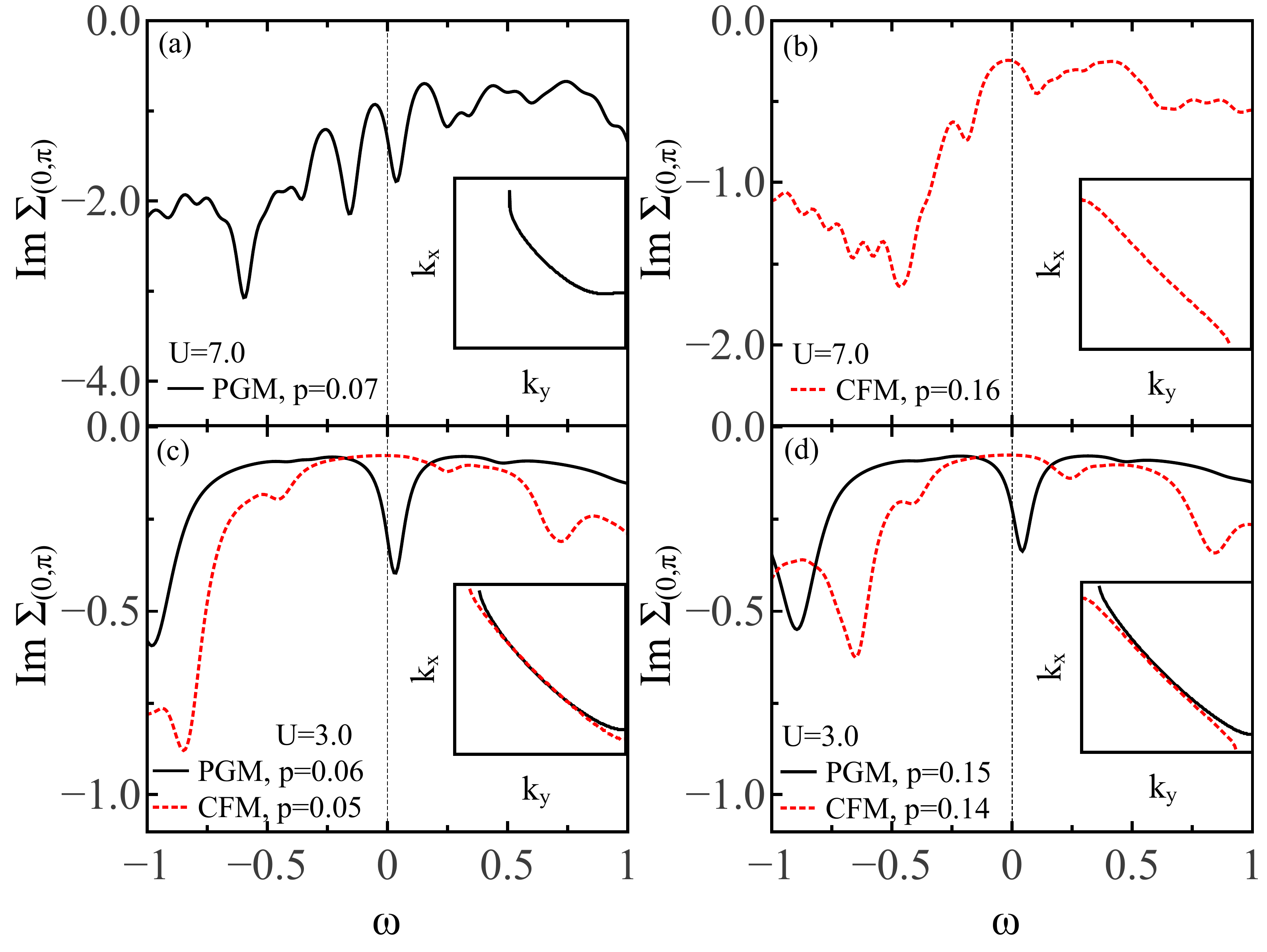} 
    \caption{Comparison of the imaginary part of the self-energy at the antinodal point $\mathbf{k}=(0,\pi)$, between CFM and PGM 
            solutions in various regions [(a) region I, (b) IV, (c) II, (d) III ] of the $U-p$ phase diagram of Fig. \ref{HF1}.
            Inset displays the corresponding FS in the first quadrant of the BZ ($k_x, k_y \in [0, \pi]$). 
%for $t'=-0.1$, $U=3.0$ and $7.0$, and different doping levels, $p$. The main panels show the imaginary part of the self-energy at the antinodal point, $\vec{k}=(0,\pi)$, calculated on the real axis. The insets display the correspondent Fermi surfaces (FSs) at the first quadrant of the Brillouin zone ($\vec{k_x}, \ \vec{k_y}=[0, \pi]$).} 
}
    \label{DOP3}
  \end{center}
\end{figure}

We confirm in the doped system the key physical properties differentiating the CFM and PGM phases, 
established for half-filling. 
Namely the PGM always breaks the FL displaying a peak in Im$\Sigma$ (see Fig.~\ref{DOP3}a,c,d), 
which has now slightly moved to positive frequency. %upon doping holes into the system. 
%This feature appears to be robust both at 
%weak [$U=3$ in Fig.~\ref{DOP3}(a) and (c)] and strong interaction [$U=7$ in Fig.~\ref{DOP3}(b)].
%, where the PGM becomes the lowest energy state.
On the other hand the CFM phase is FL-like on all the phase diagram, displaying a well behaved 
$\omega^2$ dependence of Im$\Sigma$ (Fig.~\ref{DOP3}.b,c,d).
%%%%%%%%%%%%%%%%%%%%%

\begin{figure}[h]
  \begin{center}
    \includegraphics[width=\linewidth]{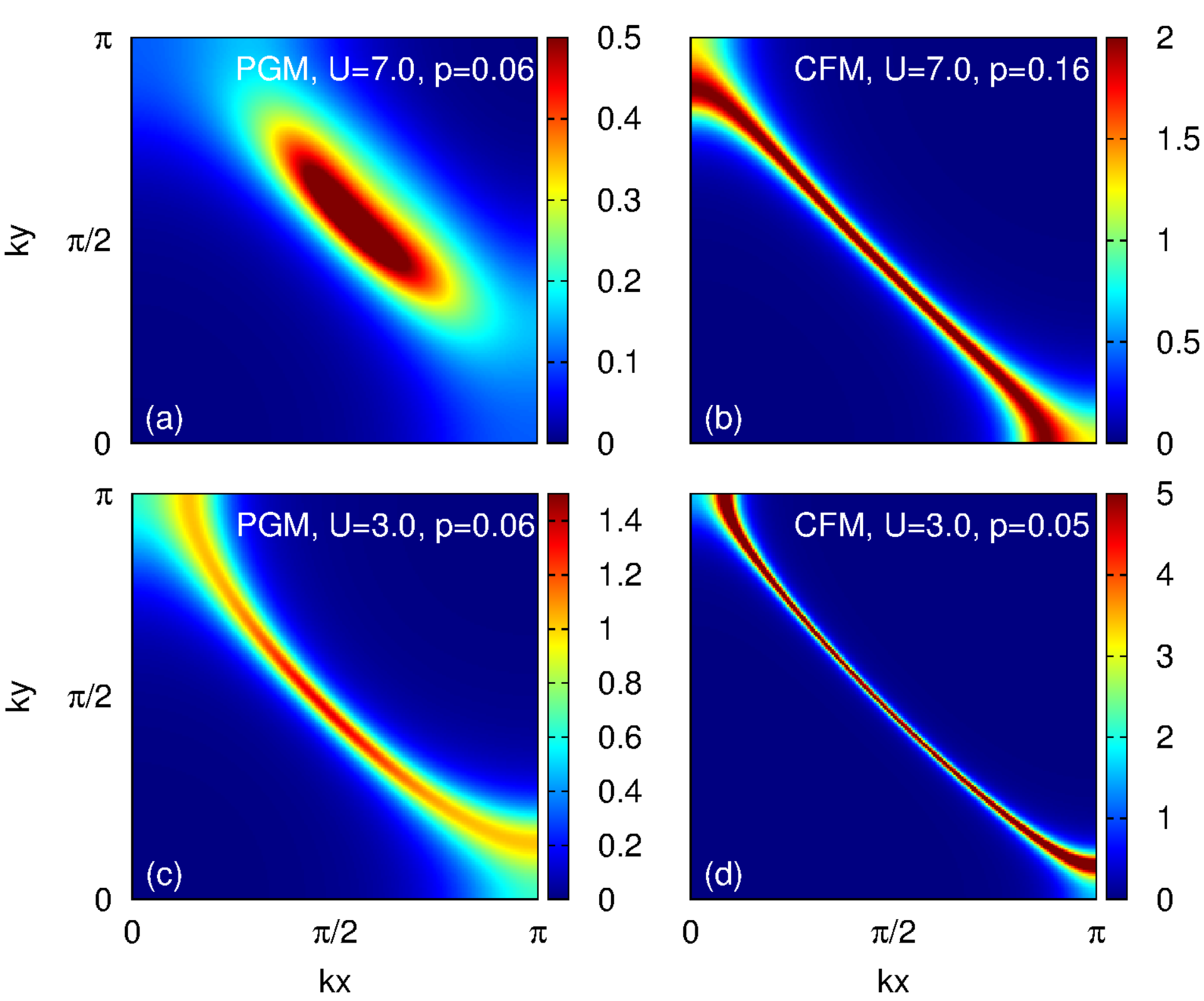}
    \caption{Spectral function $A(\mathbf{k},\omega=0)$ in the first quadrant of the BZ for the CFM and the PGM. 
             At weak interaction $U=3$ and small doping $p=0.05$, the underlying FS is hole-like 
             in both solutions. In this case however the CFM is the stable solution, not showing Fermi arcs. 
             At strong interaction $U=7.0$,
             the PGM is stable at small doping ($p=0.06$), displaying Fermi arcs and a h-FS, while 
             the CFM is stable at high doping ($p=0.16$), displaying no arc and an e-FS.} 
    \label{DOP4}
  \end{center}
\end{figure}

Let us now discuss the implications of these findings in the context of the PG phase of cuprates and its relation with the FS 
topology. This has been the subject of pioneering 
studies~\cite{maier02,senechal04,PM,YZR,liebsch09,ferrero09,sakai09,AMT10,Gull10,chen12,ferrero17}, 
though the physical mechanism at the origin of this relation 
has remained not well clarified.
%%%%%%%%%%%%%%%%%%%%%%%%%%%%%%%%%% 
The first crucial observation is that the pole in the self-energy in the PGM solution strongly enhances the scattering in the 
neighborhood of
%region of momentum space close to 
$\mathbf{k}= (0,\pi)$ (antinodes in cuprates). As a consequence, the spectral weight on the FS around $\mathbf{k}= (0,\pi)$ is strongly suppressed, giving origin
to the well known break of the FS into arcs. This can be shown in the spectral function 
$A(\mathbf{k},\omega= 0)$ plotted in Fig.~\ref{DOP4}(a),(c). %, which is obtained by using the cumulant periodization. 
These results are consistent with previous CDMFT studies~\cite{civelli05,sakai09,ferrero09} and well describe the 
angle-resolved photoemission spectroscopy measurements  on cuprates~\cite{damascelli03}. 
The CFM does not show any Fermi arc (Fig. \ref{DOP4}.b,d), rather the spectral intensity is enhanced 
at the antinodes because of the proximity to a van Hove singularity.

The second crucial observation (insets of Fig. \ref{DOP3}) is that the   
{\it PGM solution has always a h-FS}. To get some insight into this issue, 
we notice that in the PGM the low-frequency pole in the self-energy  
can be well described by $\Sigma(\mathbf{k},\omega) \simeq V^2/(\omega-\xi^{f}_\mathbf{k})$. 
This expression %is well established within CDMFT, where it 
has been derived in the framework of a low-energy model called ``hidden fermion''~\cite{Imada11,Shiro}, 
but also proposed by complementary approaches to the PG phase of cuprates~\cite{Yanaji11,YZR,P2,Morice17}.
$\xi^{f}_{\mathbf{k}}$ is located slightly above the Fermi energy in the antinodal region.
%This form is exact in $\mathbf{k}= (0,\pi)$ cluster-momentum point of the  
%PGM solutions, but it can be extended around $\mathbf{k}= (0,\pi)$ 
%(by using e.g. the cumulant periodization). 
%This is an approximation that is rather reasonable in CDMFT, especially close to the cluster-momenta points, 
%and that it has been supported by previous CT-QMC which have employed up to 4$\times$4 cluster-size (SAKAI PRB).
We see then that, if $\mathbf{k}_F$ is the Fermi wave-vector of the non-interacting system located  
on the $(0,\pi) - (\pi,\pi)$ side of the BZ, such that $\xi_{\mathbf{k}_F}=0$, the pole-like singularity of the self-energy
in the interacting system acts in such a way that the new Fermi wave-vector $\mathbf{k}'_F$ is given by 
$\xi_{\mathbf{k}'_F}-V^2/\xi^{f}_{\mathbf{k}'_F}= 0$. 
If $\xi^{f}_{\mathbf{k}'_F}$ is positive as it is for $\mathbf{k}= (0,\pi)$, it acts as an enhanced chemical potential 
(see SM section IV)
and $\mathbf{k}'_F> \mathbf{k}_F$, i.e. the interacting FS is {\it more hole-like}. 
To undergo a transition from the h-FS to e-FS in the PGM, the FS must cross the $\mathbf{k}= (0,\pi)$ point. 
But this is pre-empted by the pole-singularity of the self-energy. 
%which, providing an effective positive contribution 
%to the chemical potential, pushes the FS away from this point. 
One possibility is to have the pole singularity position
$\xi^f_{\mathbf{k}}$ move to negative frequencies. As noticed in Ref.~\cite{ferrero17}, this can be achieved by tuning $t^{\prime}$ to positive values, which is equivalent to consider electron-doped cuprates, as the same tuning can be realized by a particle-hole transformation of the Hamiltonian (\ref{HubbardModel}) that changes the sign of $t^{\prime}~$\cite{civelli05}. Another possibility is 
to have the pole singularity disappear ($V\to 0$), but in this case one loses the PGM solution. 
%This is what happens by doping the system 
%in the phase diagram of Fig. \ref{HF1}(b), where the CFM takes over the PGM at some finite doping which depends on $U$. 
We establish then an indissoluble tight relationship within the PGM solution 
between pole in the self-energy, PG, and hole-like FS. 

%%%%%%%%%%%%%%%%%%%%%%%%%%%%%%%%%%%%%%%%%%%%%%%%%%%%%%%%%%

Putting all together, we provide a rational understanding to the experimental
observations that the PG ending point is
linked to a Lifshitz transition, and above all why this appears first-order\cite{benhabib15,loret17}
or at least very sharp~\cite{piriou11,doiron17} in experiments.
Starting from weak $U$ [see Fig.~\ref{HF1}(b)] the stable solution
is the CFM, which by increasing doping presents a 
renormalized\cite{ferrero17} Fermi-liquid-like
Lifshitz transition at $p=p_{lt}$ (continuous red line
in Fig.~\ref{HF1}.b), 
where one goes continuously from a h-FS to an e-FS.
There is no PG in this case;
% (as seen in Fig.~\ref{DOP4}). 
the PG is present in the PGM, which has a h-FS (Fig. \ref{DOP4}.c), but this phase is unstable. 
When the $p_{lt}$ boundary on the $U-p$ phase diagram
meets the PGM-CFM transition boundary $p^{*}$  (black-dashed lines) at stronger
$U\simeq 5.5t$, we find for $p < p^{*}$ the PGM as the stable solution, which has a h-FS, 
while for $p > p^{*}$ the CFM is the stable solution, but this has already an e-FS
(Fig. \ref{DOP4}.b). By increasing
$p$ then the change from h-FS to e-FS is {\it bound} to the PGM-CFM phase transition
($p_{lt} \equiv p^{*}$), which is first order,
providing a correlated mechanism for the Lifshitz transition.
This is consistent with the experimental observations 
in Bi-based~\cite{piriou11,benhabib15,loret17} 
and La-based compounds\cite{doiron17}. 
% This is a new kind of correlated-driven lifshitz transition.
%This result explains why the PG of cuprates disappears rather abruptly
%in correspondance of a change of topology from a h-FS to an e-FS, 
%as observed in ref.\cite{piriou11,benhabib15,loret17,doiron17}.
On the other hand, there is a region of the phase diagram in the
range $4.5t< U< 5.5t $ where the PGM-CFM transition takes place at a doping smaller than
the Lifshitz transition of the CFM solution,
i.e. $p^{*}< p_{lt}$. At $p^{*}$ the PG disappears but the FS is
still hole-like. %, \blue{because $p^{*}$ and $p_{lt}$ are not tight yet}. 
This may account for the Tl-based~\cite{Plate05,damascelli06} and Y-based\cite{Hossain} cuprates, 
which have been reported to have a h-FS but no PG. 
This may also crucially depend on the $|t^{\prime}/t|$ value, as pointed out in Ref.\cite{ferrero17}.
%\blue{for the rather high accessible dopings.}
Our paramagnetic CDMFT phase-diagram of the two-dimensional Hubbard model can then fully account for 
apparently contradicting and debated experimental results on different members of the cuprate
family, showing that there truly exists a tight relationship between PG and FS topology. This
should manifests itself in cuprates whenever a Lifshitz transition takes place in the pseudogap phase.

%%%   \cite{doiron17}
%\blue{The existence of these experimental results and the presence of the small region in the phase-diagram in which $p^{*}$ and $p_{lt}$ are not tight yet, may indicate that the relation between the first order phase transition between both metal and the Lifshitz transition may not be so general in the curates physics. On the other hand, it is very interesting that the CDMFT description of the two dimension Hubbard model is able to accounts for different experimental results (with different compounds), in different regions of the phase diagram.}

%%%%%
% 
In conclusion, we have studied the paramagnetic normal-state of the two-dimensional Hubbard model at
zero temperature for a broad
range of interaction, $U$, frustration, $t^{\prime}$, and doping level, $p$.
Our main finding is the coexistence of a correlated Fermi liquid metal (CFM) with a
non-Fermi liquid metal (PGM). At half-filling, we answer to an open debate
by showing that the CFM is the stable solution and displays 
a correlation-driven Mott MIT, differently from the PGM which displays instead a PG in the spectra.
Next we show that for strong interactions and small doping, region relevant for underdoped cuprates, the PGM becomes the stable solution. %displaying always a h-FS. 
This result is at the origin of a correlation-driven Lifshitz transition, 
as by increasing doping a first-order transition takes place 
from the PGM phase, which has a h-FS, to the CFM, which at this interaction values has an e-FS.
Our theory rationalizes the variety displayed on the phase diagram of the cuprate family,
telling us that if the PG meets a Lifshitz transition, then it should collapse. 
This implies also that the PG cannot exists on an e-FS.
% 
%recent experiments, 
% and has always a hole-like Fermi surface
%(h-FS). The PGM is unstable with respect to the CFM, having a higher
%energy, except for strong interactions and small doping region, which is relevant for underdoped cuprates
%materials. Increasing doping, one finds the CFM, which has an electron-like Fermi surface (e-FS). 
%\gr{[SS: Do we need to CFM, PGM, h-FS,e-FS again? It is fine if you think this is helpful for readers who read only conclusion...]}
%A doping-driven first-order transition from the PGM to the CFM is therefore always accompanied
%by a change of topology from h-FS to e-FS, unveiling a correlated-driven Lifshitz transition mechanism and 
%Our theory gives a rational phase diagram, where exceptions to this rule (like the
%Tl-based and Y-based compounds) could be accounted for.
%Our results should also stimulate further investigations on 
%the origin of the PGM,
%whose behavior at half-filling 
%and whose high kinetic energy 
%The PGM 
%is reminiscent of Anderson's RVB theory, though relevant differences
%have been already pointed out~\cite{Shiro}. 
The behavior of the PG that is sensitive to the FS topology
must be contrasted with the one of superconductivity, which does not appear 
much affected by the Lifshitz transition\cite{loret17}.
This may represent the key to finally
unveil the true nature of the relation between the PG and 
the high-$T_c$ mechanism. Future experimental and
theoretical investigations should be pursued in this direction.

%\gout{The relationship of the PGM and}
%\blue{ 
%\gout{superconductivity could reveal the high-T$_C$ mechanism,
%but this link has been missing so far.} 
During our investigations, we became
aware of the work of Ref.~\cite{ferrero17}, whose results are in good
agreement with ours. Our PGM, however, is found as a
second metastable solution distinct from the CFM. 
%one
%on a wide part of the doping-interaction phase diagram. 
This in particular discloses the origin of the tight link between the PG and 
correlated first-order Lifshitz transition.

We thank I. Paul for his insightful comments.
We acknowledge discussion with W. Wu, M. Ferrero, A. Georges, B. Loret, A. Sacuto, O. Parcollet,
A.-M. Tremblay, G. Sordi. 
This  work  was  supported  by CNPq, CAPES and FAPEMIG, and grants JSPS KAKENHI (16H06345 and 17K14350), support from
the INCT on Quantum Information/CNPq is also gratefully acknowledged. 
Part of the results were obtained at CENAPAD-SP.

\clearpage

\setcounter{equation}{0}
\setcounter{figure}{0}
\setcounter{table}{0}
\setcounter{page}{1}

\onecolumngrid

\begin{center}

 {\large \textbf{Supplementary Material for \\ ``Correlation-driven Lifshitz transition at the emergence of the pseudogap phase in the two-dimensional Hubbard Model''}}
 
 \normalsize
 
 \vspace{0.5cm}
 
 Helena Bragan\c{c}a$^{1,2}$, Shiro Sakai$^3$, M. C. O. Aguiar$^1$, Marcello Civelli$^2$\\
 
 \vspace{0.5cm}
 
 {\small{\textit{$^1$Departamento de F\'isica, Universidade Federal de Minas Gerais,
C. P. 702, 30123-970, Belo Horizonte, MG, Brazil\\
$^2$Laboratoire  de  Physique  des  Solides,  Univ.   Paris-Sud,
Universit\'e  Paris-Saclay,  CNRS  UMR  8502,  F-91405  Orsay  Cedex,  France\\
$^3$Center for Emergent Matter Science, RIKEN, Wako, Saitama 351-0198, Japan}}}

\normalsize

 \end{center}

 \vspace{0.5cm}

\twocolumngrid

\thispagestyle{empty}
   
%%%%%%%%%%%%%%%%%%%%%%%%%%%%%%%%%%%%%%%%%%%%%%%%%%%%%%%%%%%%%%%%%%%%%%%%%%%%%%%%%%%%%%%%%%%%%%%%%%%%%%%%%%%%%%%%%%%%%%%%%%%%%%

\section{I - Bath description within ED-CDMFT}

Within cellular dynamical mean field theory the Hubbard model (Eq.~\ref{HubbardModel}) is mapped onto a cluster Anderson model with $2\times2$ interacting sites surrounded by a non-interacting bath. The latter Hamiltonian is given by
\begin{eqnarray}
H&=&\sum_{i j}^{N_c} \sum_{\sigma }E_{i j \sigma } c^{\dagger}_{i \sigma} c_{j \sigma} + U \sum_{i}^{N_c} c^{\dagger}_{i \uparrow} c_{i \uparrow} c^{\dagger}_{i \downarrow}c_{i \downarrow} \nonumber \\ &+& \sum_k \sum_{\sigma} \varepsilon_{k \sigma}a^{\dagger}_{k \sigma} a_{k \sigma} + \sum_l^{N_b} \sum_{i \sigma} \left( V_{l i \sigma} a^{\dagger}_{l \sigma} c_{i \sigma} + h.c.\right).
\label{AndCluster}
\end{eqnarray}
$N_c=4$ for the $2\times2$ plaquette and the indexes $i$ and $j$ label the cluster sites.  $c^{\dagger}_{i \sigma}$ and $a^{\dagger}_{l \sigma}$ create electrons with spin $\sigma$ on the cluster and the bath, respectively.  
The matrix elements $E_{ i j \sigma}$ represent the hopping parameters of the original Hubbard model (i.e., $E_{ii}=\mu$, $E_{i,i+1}=t$, $E_{i,i+2}=t'$) while  
the bath dispersion $\varepsilon_{k\sigma}$ and the cluster-bath hybridization $V_{l i \sigma}$ are self-consistently determined. 

To solve the above Hamiltonian with exact diagonalization one have to truncate the bath to a finite number of sites, although the original Hubbard model is still in the thermodynamic limit. In this work we use $N_b=8$ bath sites, separated in two sub-baths with the same geometry of the cluster.
In this case, the Anderson Hamiltonian can be represented as 
\begin{eqnarray}
H&=&\sum_{\sigma}\Psi^{\dagger}_{\sigma}E_{\sigma}\Psi_{\sigma}+U \sum_{i}n_{i \uparrow}n_{i,\downarrow} + \nonumber \\ &&+ \sum_{\alpha \sigma} \left( \Phi^{\dagger}_{\alpha \sigma} E_{B \sigma}^{\alpha} \Phi_{\alpha \sigma} + 
\Phi^{\dagger}_{\alpha \sigma}V_{\alpha \sigma} \Psi_{\sigma} + h.c. \right), 
\end{eqnarray}
where $\alpha =1,2$ labels the sub-bath, $E_{\sigma}$, $E_{B \sigma}^{\alpha}$ and 
$V_{\alpha \sigma}$ are $4 \times 4$ matrices, $\Psi^{\dagger}_{\sigma}=(c^{\dagger}_{1\sigma}, c^{\dagger}_{2\sigma}, c^{\dagger}_{3\sigma},c^{\dagger}_{4\sigma})$ and $\Phi^{\dagger}_{\sigma}=(a^{\dagger}_{1\sigma}, a^{\dagger}_{2\sigma}, a^{\dagger}_{3\sigma},a^{\dagger}_{4\sigma})$.

In the most general form (relaxed bath parametrization), the bath parameters are given by
\begin{equation}
E_{B \sigma}^{\alpha}=  \begin{pmatrix}
    \varepsilon^{\alpha}_{1 \sigma} & 0 & 0 & 0 \\
    0 & \varepsilon^{\alpha}_{2 \sigma} & 0 & 0 \\
    0 & 0 & \varepsilon^{\alpha}_{3 \sigma} & 0\\
    0 & 0 & 0 & \varepsilon^{\alpha}_{4 \sigma} \\
  \end{pmatrix}
  \label{P1-1}
\end{equation}
and

\begin{equation}
V_{\alpha \sigma}=  \begin{pmatrix}
    V^{\alpha}_{1 1 \sigma} & V^{\alpha}_{1 2 \sigma} & V^{\alpha}_{1 3 \sigma} & V^{\alpha}_{1 4 \sigma} \\
    V^{\alpha}_{2 1 \sigma} & V^{\alpha}_{2 2 \sigma} & V^{\alpha}_{2 3 \sigma} & V^{\alpha}_{2 4 \sigma} \\
    V^{\alpha}_{3 1 \sigma} & V^{\alpha}_{3 2 \sigma} & V^{\alpha}_{3 3 \sigma} & V^{\alpha}_{3 4 \sigma} \\
    V^{\alpha}_{4 1 \sigma} & V^{\alpha}_{4 2 \sigma} & V^{\alpha}_{4 3 \sigma} & V^{\alpha}_{4 4 \sigma} \\
  \end{pmatrix}.
\label{P1-2}
  \end{equation}

Within this representation, one have $2*\left(N_b +N_c*N_b\right)$ free bath parameters to determine throughout the self-consistent calculation (the factor 2 can be suppressed in the paramagnetic case, since $E_{B \uparrow}^{\alpha}=E_{B \downarrow}^{\alpha}$ and  $V_{\alpha \uparrow}=V_{\alpha \downarrow}$). Alternatively, one can use a more efficient (constrained) bath parametrization~\cite{GK2,Marcello2009SM}, which is faster and simpler to interpret:

\begin{equation}
E^{\prime \alpha}_{B \sigma}=  \begin{pmatrix}
    \varepsilon^{\alpha}_{\sigma} & t^{\alpha}_{B \sigma} & t'^{\alpha}_{B \sigma} & t^{\alpha}_{B \sigma} \\
    t^{\alpha}_{B \sigma} & \varepsilon^{\alpha}_{ \sigma} & t^{\alpha}_{B \sigma} & t'^{\alpha}_{B \sigma} \\
    t'^{\alpha}_{B \sigma} & t^{\alpha}_{B \sigma} & \varepsilon^{\alpha}_{ \sigma} & t^{\alpha}_{B \sigma}\\
    t^{\alpha}_{B \sigma} & t'^{\alpha}_{B \sigma} & t^{\alpha}_{B \sigma} & \varepsilon^{\alpha}_{ \sigma} \\
  \end{pmatrix}
  \label{P2-1}
\end{equation}
and

\begin{equation}
V'_{\alpha \sigma}=  \begin{pmatrix}
    V^{\alpha}_{ \sigma} & 0 & 0 & 0 \\
    0 & V^{\alpha}_{\sigma} & 0 & 0 \\
    0 & 0 & V^{\alpha}_{ \sigma} & 0 \\
    0 & 0 & 0 & V^{\alpha}_{\sigma} \\
  \end{pmatrix}.
  \label{P2-2}
\end{equation}

In the latter representation, we have explored the translational symmetry of the bath, considering that all the bath sites inside a given sub-bath have the same energy, and also that each cluster site hybridizes only with the correspondent site of the plaquette sub-bath. To compensate the simplification, we have included hopping between nearest-neighbor and next-nearest-neighbor sites within each sub-bath. In the simplified parametrization, we only have $2*\left(4*M_b \right)$ parameters to determine self-consistently, where $M_b=N_b/N_c$ is the number of sub-baths.   

The parametrizations are related through the unitary matrix $S$ ($SS^{T}=1\!\!1$) which diagonalizes $E^{\prime \alpha}_{B \sigma}$, that is, 
$E_{B \sigma}^{\alpha}=SE^{\prime \alpha}_{B \sigma }S^{T}$ and $V_{\alpha,\sigma}=S^{T}V'_{\alpha \sigma}$, so that the bath function $
 \Delta (i \omega)= V^{T}\left( i \omega 1\!\!1 - E_B \right)^{-1} V$ remains invariant over the bath transformation. 

Throughout this work, we have used the reduced bath parametrization to produce the phase diagrams displayed in Fig.~\ref{HF1}. For some range of parameters, however, we have relaxed the bath parameters to confirm that our conclusions about the existence of different solutions and the stability of each one in different regions of the phase-diagram do not depend on the choose of the parametrization. 

As an example, the converged bath parameters for the four relevant cases displayed at Fig.~\ref{DOP4} can be seen on Table~\ref{Tab:BP}. Note that the results do not depend on the spin, since we concentrate on the paramagnetic solution. The eigenvalues of the diagonalized $E^{\prime \alpha}_{B \sigma}$ matrix are given by $\{ \varepsilon^{\alpha} - t_B'^{\alpha}, \ \varepsilon^{\alpha} - t_B'^{\alpha}, \ \varepsilon^{\alpha} -2t_B^{\alpha} + t_B'^{\alpha}, \ \varepsilon^{\alpha}+2t_B^{\alpha} + t_B'^{\alpha}\}$. For the four cases displayed on the table, we have the following set of eigenvalues: PGM, U=3.0, p=0.06 $\{0.0768, \ 0.0768, \ -0.1212, \ 0.2548\}$; CFM, U=3.0, p=0.05 $\{-0.039, \ -0.039, \ -0.033, \ 0.031\}$; PGM, U=7.0, p=0.06 $\{0.074, \ 0.074, \ 0.052, \ 0.180 \}$; CFM, U=7.0, p=0.16 $\{-0.038, \ -0.038, \ -0.046, \ 0.03\}$.

\begin{table}[h]
 \begin{center}
\includegraphics[width=\linewidth]{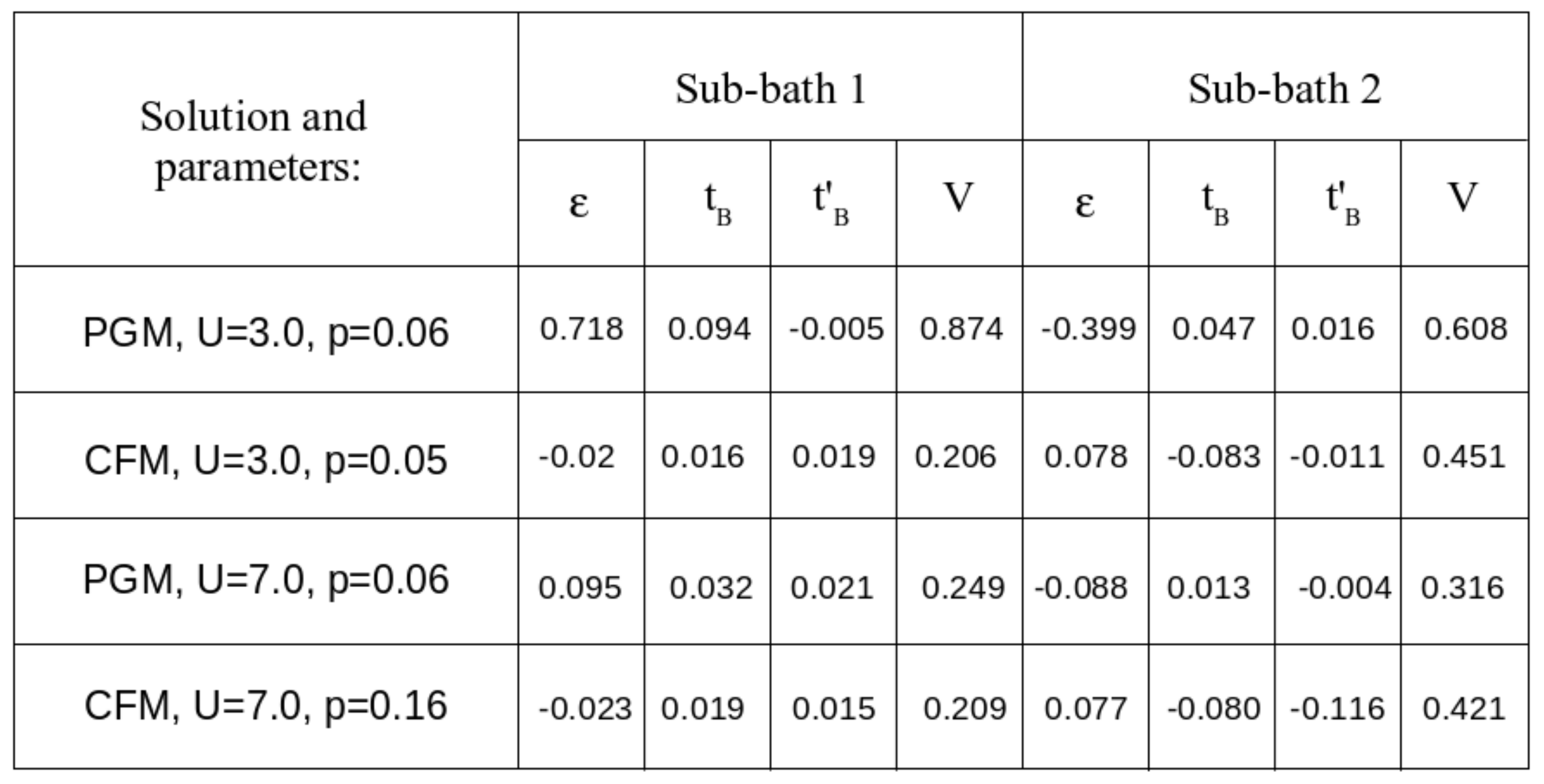}
  \caption{Converged bath parameters for the auxiliary Anderson model within the constrained parametrization (Eqs.~\ref{P2-1} and~\ref{P2-2}).}
  \label{Tab:BP}
 \end{center}
\end{table}

\begin{figure}[h]
 \begin{center}
\includegraphics[width=\linewidth]{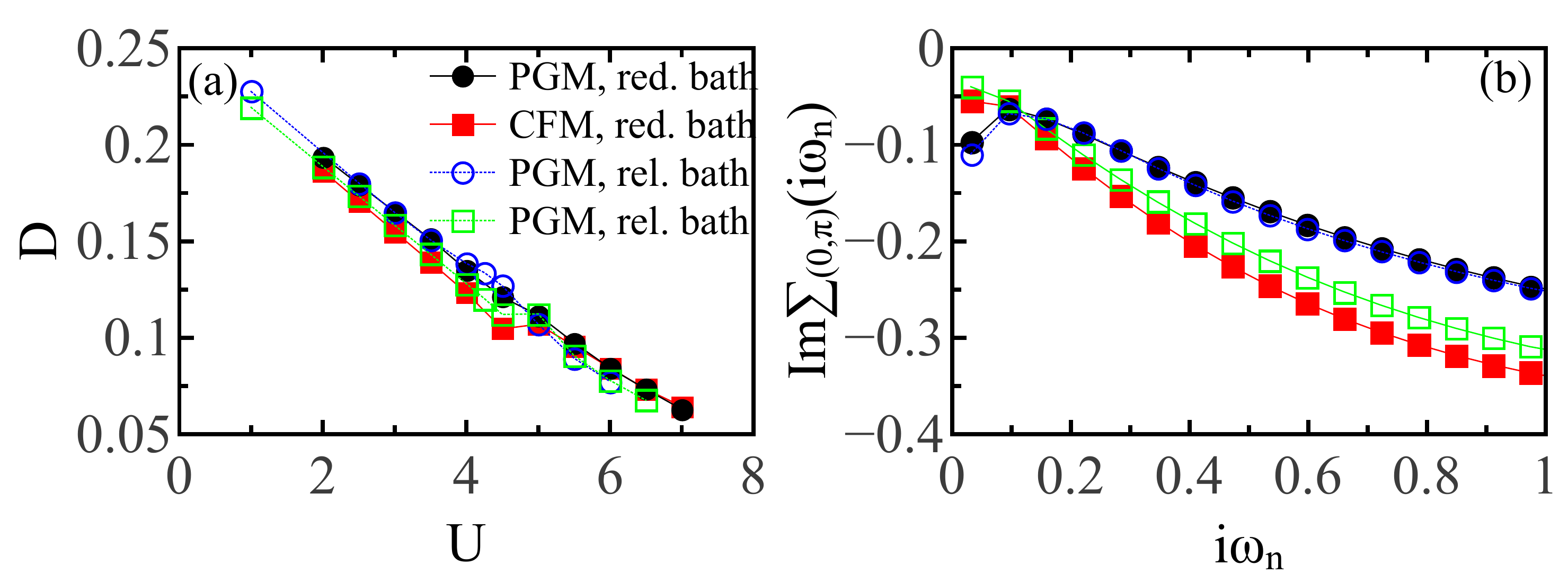}
  \caption{Comparison between results obtained with different bath parametrization for the half-filled case with $t'=0$. (a) Double occupation probability and (b) imaginary part of the self-energy at the anti-nodal point, calculated on the imaginary axis for $U=3.0$. Open symbols correspond to the relaxed bath parametrization (Rel. bath), while filled symbols were obtained with the constrained parametrization (Red. bath). With both bath parametrizations, we can distinguish two different solutions, a PGM (circles) and a CFM (squares). }
  \label{Relax}
 \end{center}
\end{figure}

A comparison between results obtained with the reduced (Eqs.~\ref{P2-1} and~\ref{P2-2}) and relaxed (Eqs.~\ref{P1-1} and~\ref{P1-2}) bath parametrizations can be seen on Fig.~\ref{Relax}, which shows the double occupation probability and the imaginary part of the self-energy at the anti-nodal point, calculated on the imaginary axis. The results are for the half-filled case, with $t'=0$. With both parametrizations, we can clearly distinguish the different solutions, CFM and PGM.   

Under the numerical calculation, the bath-parameters $\varepsilon^{\alpha}$, $t_B^{\alpha}$, $t_B'^{\alpha}$ and $V_{\alpha}$ are determined at each CDMFT-iteration by fitting the Anderson-impurity Weiss field with a $N_b$-pole bath function $\hat{\mathcal{G}}_{N_b}^{new}(i \omega_n)= i \omega_n 1\!\!1 - \hat{E} - \hat{\Delta}$, with  $
 \Delta (i \omega)= V^{T}\left( i \omega 1\!\!1 - E_B \right)^{-1} V$. The fitting is obtained via a conjugate gradient minimization algorithm, with a distance function that emphasizes the lowest frequencies~\cite{Marcello2009SM, Capone2004SM} 
 
 \begin{equation}
f=\sum_n \sum_{ij} \frac{1}{\omega_n}|\mathcal{G}_0^{new}(\omega_n)-\mathcal{G}_{N_b}^{new }(\omega_n)|_{ij}. 
\end{equation}

The convergence criteria combines a convergence of the cluster Green's function between consecutive iterations with a small value of the distance function.  
%%%%%%%%%%%%%%%%%%%%%%%%%%%%%%%%%%%%%%%%%%%%%%%%%%%%%%%%%%%%%%%%%%%%%%%%%%%%%%%%%%%%%%%%%%%%%%%%%%%%%%%%%%%%%%%%%%%%%%%%%%%%%%

\section{II - Lattice quantities in momentum space and energy calculation}
 
 As discussed in the main text, the cellular dynamical mean field calculation in a 2$\times$2 plaquette cluster gives us frequency dependent quantities in specially symmetric points of the first quadrant of the Brillouin zone, $\vec{k}=(0, \pi), \ (\pi, 0), \ (\pi,\pi)$ and $(0,0)$. 
 The reconstruction of the lattice quantities in a broader range of momentum can be obtained through a (truncated) Fourier expansion in which the cluster quantities are the expansion coefficients, that is, 

\begin{equation}
 Q^L(k,i\omega_n)= \frac{1}{N_c} \sum_{i,j=1}^{N_c} Q^c_{ij}(i\omega_n)\mbox{exp}[i \vec{k} \cdot (\vec{r}_i-\vec{r}_j)],
\end{equation}
where the sub-indices $L$ and $c$ stand for lattice and cluster quantities, respectively, and  $\vec{r}_i$ and $\vec{r}_j$ are spacial intra-cluster coordinates. 

Different periodization schemes have been proposed in the literature, related to different choices for the quantity $Q$, such as the self-energy scheme, $\Sigma$~\cite{Marcello2005}, the cumulant scheme, $M=\left[i \omega_n+ \mu -\Sigma\right]^{-1}$~\cite{Tudor06,Tudor}, and the Green's function one~\cite{GK1,GK2}. 

A comparison between results obtained with the three different procedures can be found in reference~\cite{TeseMarcelloSM}; it is known that the choices $Q=M$ and $Q=G$ are equivalent in first order expansion and produce the best and the least cluster-size-dependent results close to the Mott transition~\cite{SIZESM}. 
For this reason, we use the $M-scheme$ throughout this work.  

In this case, we obtain the lattice cumulant from the cluster ones; the lattice self-energy and the lattice Green's function can then be obtained through the relations

\begin{equation}
 G^L(k, i \omega_n ) = \left[M^L(k,i \omega_n)^{-1}- \varepsilon(k)\right]^{-1}
\end{equation}
where $\varepsilon(k)= -2t (\cos k_x + \cos k_y)- 4t^{\prime} \cos k_x \cos k_y$
and
\begin{equation}
 \Sigma^L(k, i \omega_n ) = i \omega_n + \mu -M^L(k,i \omega_n)^{-1}.
\end{equation}

 \begin{figure}[h]
  \begin{center}
    \includegraphics[width=0.9\linewidth]{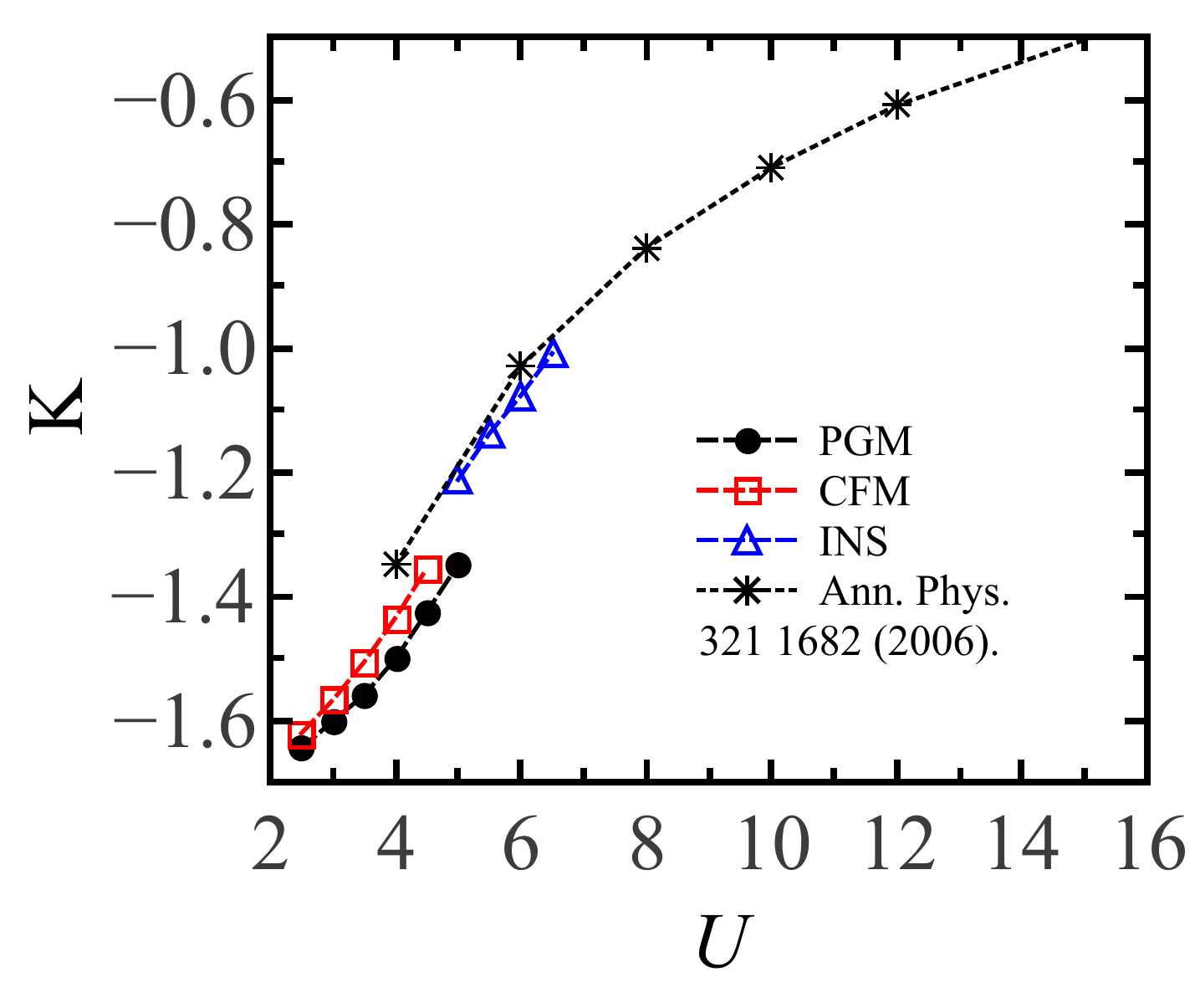}
    \caption{Kinetic energy as a function of $U$ for half-filling and $t^{\prime}=0.0$. Comparison between our results and the one shown in reference \cite{Tudor}. } 
    \label{Ap2}
  \end{center}
\end{figure}

Once we have $G^L(k, i \omega_n)$, we can calculate the kinetic energy with the relation
\begin{equation}
 K=\frac{1}{N_k \beta}\sum_{\sigma, k, n } \xi_{k} G(k,i \omega_n), 
\end{equation}
where $\xi_{k}= \, \varepsilon_{k}- \mu$ is the electronic dispersion for the two-dimensional square lattice, $N_k$ is the number of points in momentum space and $\beta = 100/ t$ is the inverse effective temperature 
(used as a Matsubara grid in the ED-CDMFT implementation). For the paramagnetic solution, the summation over spin corresponds to a factor 2. The summation over the Matsubara frequency, on the other hand, is more complicated and have to be done carefully in order to properly account for the tails for large frequency~\cite{MahanSM}.  

A comparison between our calculation of the kinetic energy and other present in the literature can be seen on Fig.~\ref{Ap2}, which shows $K$ as a function of $U$ for half-filling, $t^{\prime}=0.0$, and our three different solutions, that is, the CFM, PGM and Mott insulator. 

The potential energy, on the other hand, is calculated through $U \, D$, where $D=\sum_i \langle n_{i\uparrow}n_{i\downarrow}\rangle$ is the double occupation probability.  

\begin{figure}[h]
  \begin{center}
    \includegraphics[width=\linewidth]{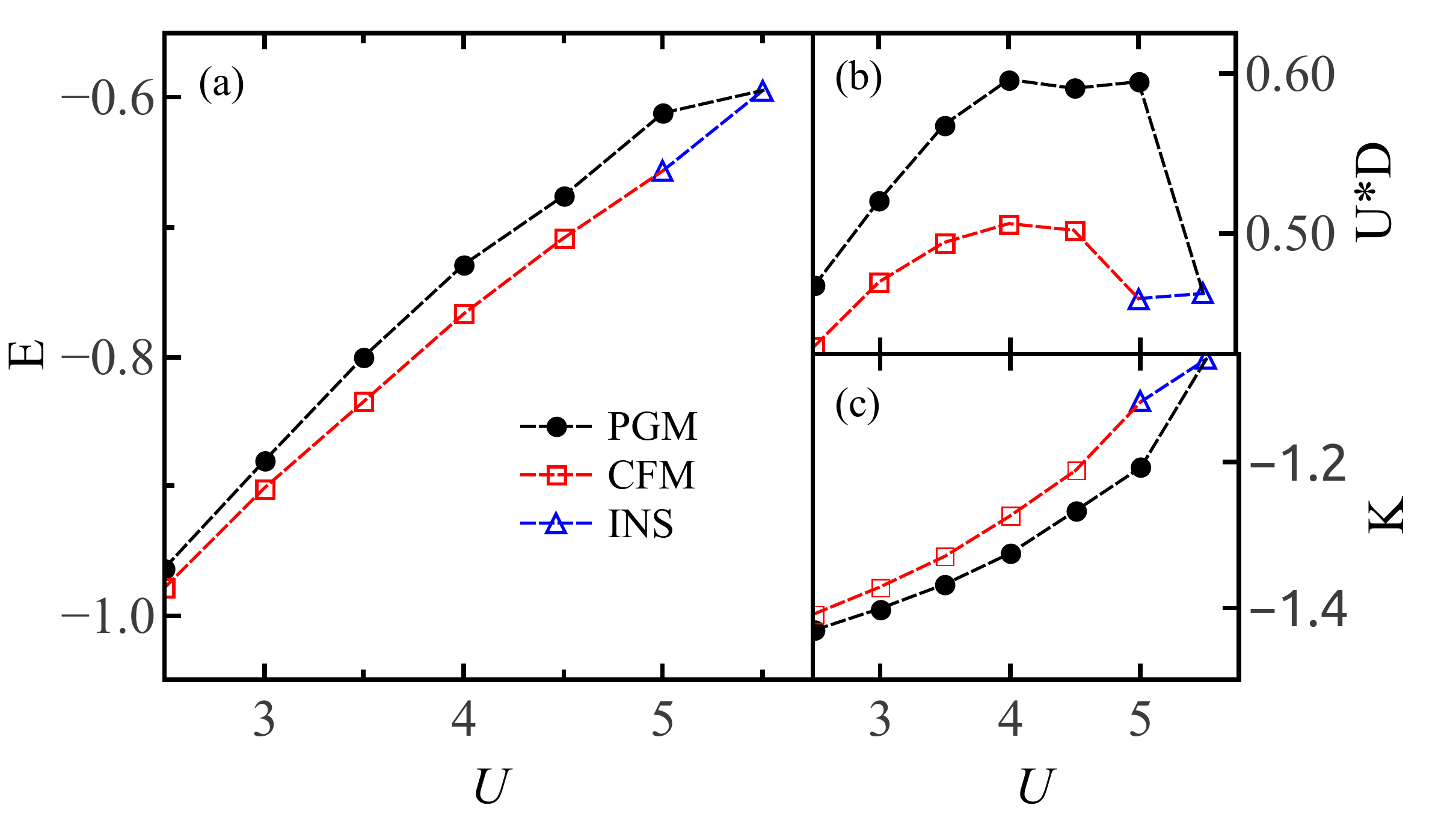}
    \includegraphics[width=\linewidth]{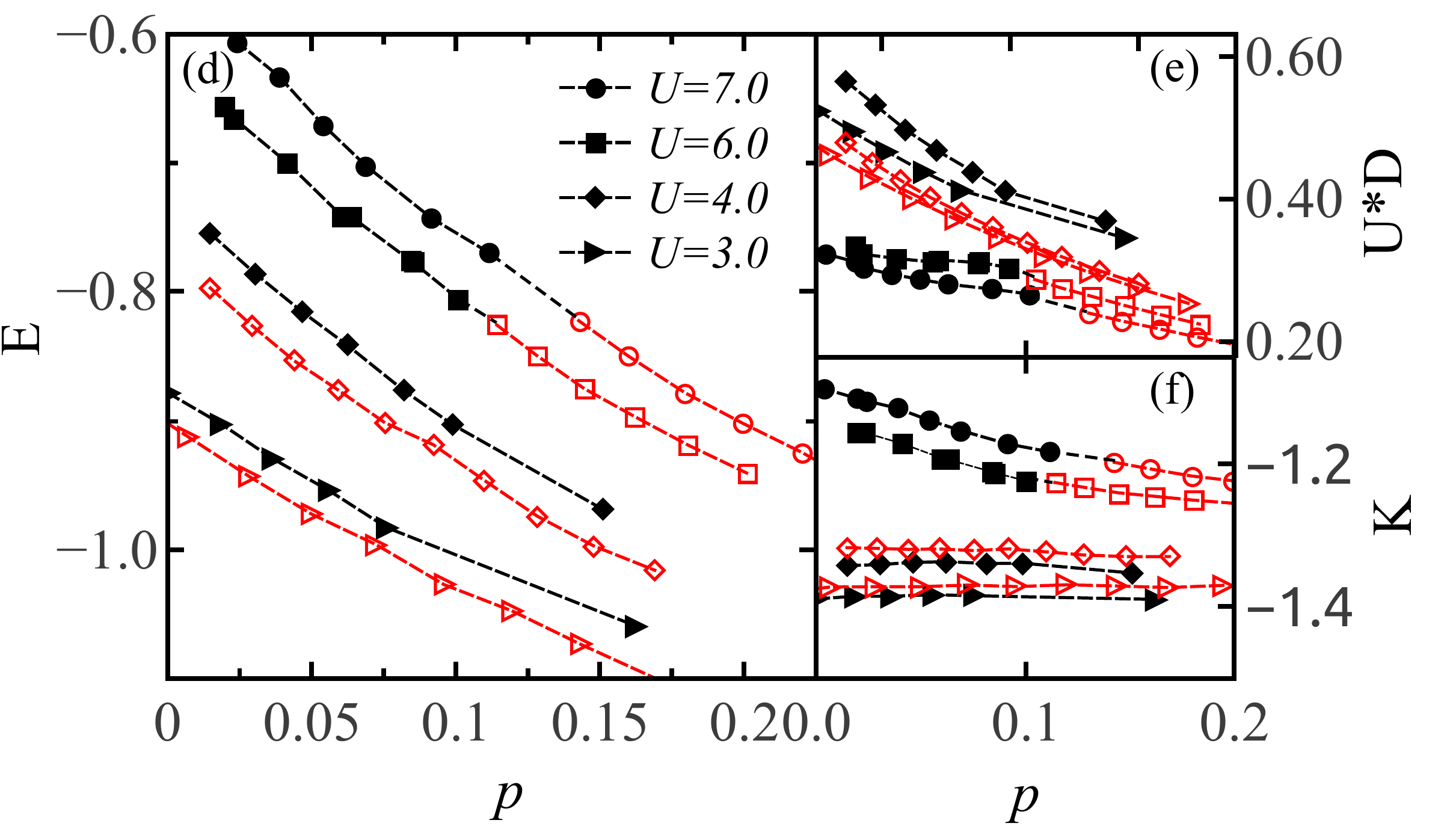}
    \caption{Total, potential and kinetic energies (a)-(c) as a function of interaction at half-filling, and (d)-(e) as a function of doping for different interaction strengths. All the results were obtained with $t^{\prime}=-0.1$. Red-open symbols represent the CFM solution, while black-filled ones correspond to the PGM phase.  } 
    \label{Ap3}
  \end{center}
\end{figure}

The total energy $E= K+ U \,D$, the potential ($U \, D$), and the kinetic ($K$) energies for the different solutions are displayed in Fig. \ref{Ap3}. Panels (a)-(c) show the energy as a function of interaction for the half-filled case; panels (d)-(e) correspond to the doped case, for different values of interaction. 

As discussed in the main text, when both CFM and PGM solution are present, the CFM is the stable solution (lower total energy). 
It is useful to separate the potential and kinetic energy contributions to the total energy. 
We observe that the PGM has a higher double occupation of sites $D$, i.e. a higher potential energy $U D$, 
and a lower (negative) kinetic energy $K$ than the CFM. 
This is reminiscent of the Anderson's Resonating Valence Bond solution to the Heisenberg model~\cite{Anderson97-SM}, 
which favors a gain in kinetic energy with respect to the paramagnetic solution~\cite{TocchioSM}. 

 \section{III - Spectral function in a broader frequency range}

 \begin{figure}[h]
  \begin{center}
    \includegraphics[width=\linewidth]{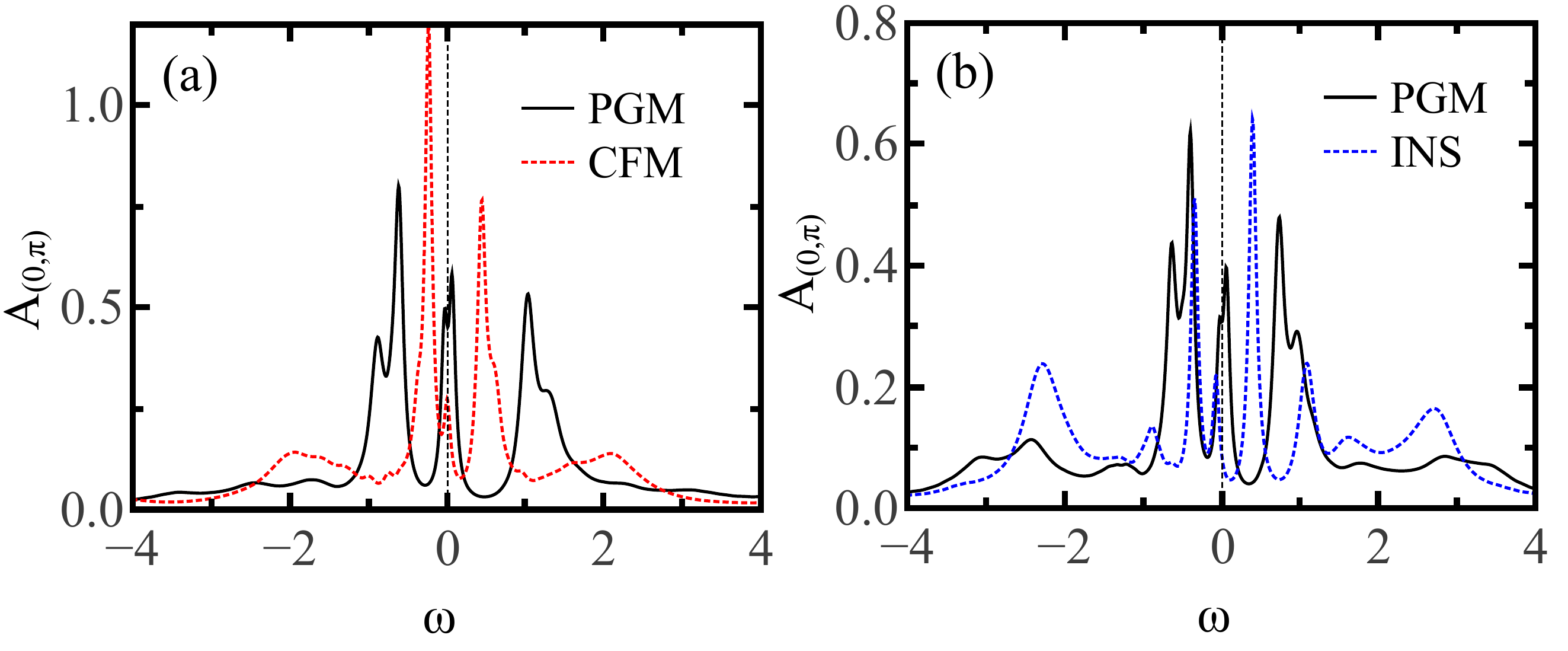}
    \caption{ Spectral function $A(\mathbf{k},\omega)$ in the $\mathbf{k}=(0,\pi)$ point of momentum space 
      for half-filling, $t^{\prime}=-0.1$, and 
     different co-existing solutions, (a) CFM and PGM, for $U=4.0$ and (b) INS and PGM, for $U=5.0$.} 
    \label{Ap1}
  \end{center}
\end{figure}
 
Fig. \ref{Ap1} shows the spectral function $A(\omega)_{(0,\pi)}$ at the antinodal point $\mathbf{k}= (0,\pi)$, 
at half-filling and $t^{\prime}=-0.1$, for $U=4.0$ and $U=5.0$ respectively. For $U=4.0$ one can observe a 
co-existence between a correlated Fermi metal and a pseudogap metal. For $U=5.0$, we find the pseudogap metal 
and a Mott insulator close to metal transition point $U_c$, 
displaying then a small gap. These are the same data shown on Fig. 2 of the main text, but with a larger frequency range.
The Hubbard sub-bands are clearly visible in the insulator around $\omega=|U/2|$ (Fig. \ref{Ap1}.b), but their formation is
already observable in the CFM (Fig. \ref{Ap1}.a) and the PGM (Fig. \ref{Ap1}.b).      
 
\section{IV - Renormalized energies and Fermi surface topology}

\begin{figure}[h]
  \begin{center}
    \includegraphics[width=\linewidth]{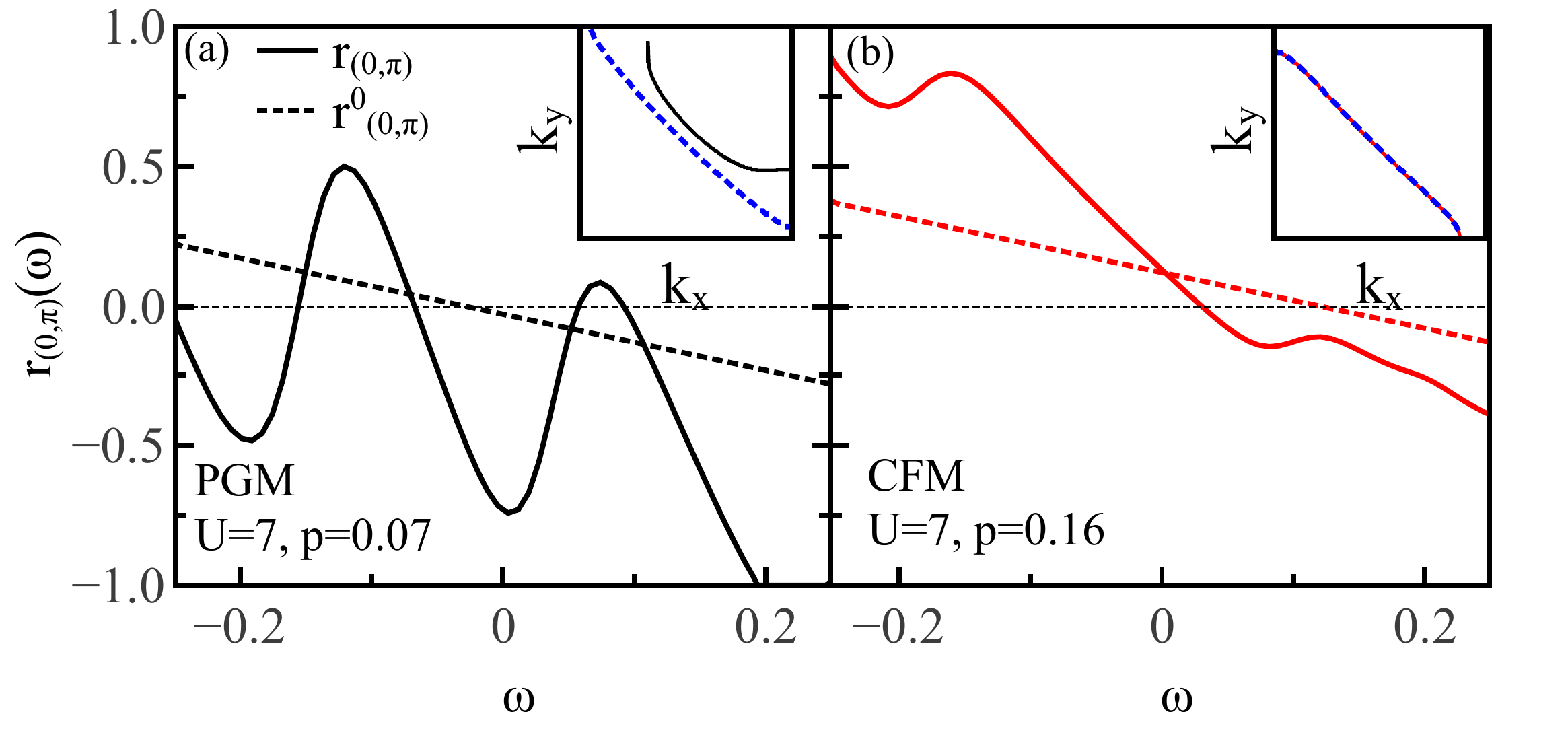}
    \caption{Renormalized energies, $r_{(0,\pi)}(\omega)$ and $r^0_{(0,pi)}(\omega)$, at low frequency, 
    calculated at the antinodal point, $\vec{k}=(0, \pi)$, for $t'=-0.1$. Insets:
    Interacting Fermi surface (continuous line) compared with the non-interacting one (dashed line).} 
    \label{Ap4}
  \end{center}
\end{figure}

In order to see the effect of the correlation on the Fermi surface topology,
one can consider the renormalized energy \cite{Tudor}:
\begin{equation}
r_k(\omega)=\,-\text{Re} G(k,\omega)^{-1} = 
\text{Re}\Sigma_{k}(\omega)-\omega+ \varepsilon_{k} -\mu.
\end{equation}
For a given $k$ in momentum space, the $k$-state is occupied if $r_k(0)<0$.
If we consider then the cluster momentum $k= (0,\pi)$, a metallic solution 
has a hole-like Fermi surface if $r_{(0, \pi)}(0)<0$, while it has an electron-like 
Fermi surface if  $r_{(0, \pi)}(0)>0$ (i.e. the $k= (0,\pi)$-state is empty)

In Fig. \ref{Ap4} we display $r_{(0, \pi)}(\omega)$ and compare it with 
$r_{(0, \pi)}^0(\omega)= -\omega+ \epsilon_{(0, \pi)} -\mu_0$,
where $\mu_0$ is the chemical potential of the non-interacting 
system ($U= 0$) which gives the same particle density. 

Panel (a) shows the PGM for $U= 7t$ and rather small doping $p=0.07$,
where the solution is the stable one. With respect to the 
$U=0$, one can observe that there is the pole contribution to the self-energy
close to $\omega=0$:
$$\text{Re}\Sigma_{k}(\omega)\simeq \mu_0+ \frac{V^2}{\omega-\xi_k^f}$$
with $\xi_k^f >0$, which pulls down $r_k(\omega=0)$ to more negative 
values than the non-interacting one $r_k^0(\omega=0$). Therefore
if one starts with a non-interacting Fermi surface at $U=0$ which is hole-like, 
and turns on $U$, the resulting interacting Fermi surface is even more hole-like  
(see inset of Fig.\ref{Ap4}.a).
If one consider the Fermi-liquid CFM solution instead, displayed for 
$U= 7t$ and rather high doping $p=0.16$ where it is stable (panel b), 
the self-energy has no pole at low frequency, hence the effect of 
the interaction on the Fermi surface shape are much milder. 
In fact, for $p=0.16$, if one starts
an electron-like Fermi surface for $U=0$ ($r_k^0(\omega=0)>0$), 
and then turns the interaction on,
the resulting interacting Fermi surface remains electron-like 
(also $r_k(\omega=0)>0$). In this case the difference between 
the non-interacting and interacting Fermi surfaces is barely 
visible.

\end{document}